w% ****** Start of file apssamp.tex ******
\documentclass[showpacs,printnumbers,amsmath,amssymb,twocolumn]{revtex4}

\usepackage{graphicx}% Include figure files
\usepackage{dcolumn}% Align table columns on decimal point
\usepackage{bm}% bold math
\usepackage{xspace} % defines the xspace command for use at the end of macros that produce text

\newcommand{\ket}[1]{|{#1}\rangle}
\def\Rb87{$^{87}\text{Rb}$}
\def\0{\ket{0}}
\def\1{\ket{1}}
\begin{document}

%\preprint{APS/123-QED}

\title{Sympathetic Wigner function tomography of a dark trapped ion}

\author{Safoura Sadat Mirkhalaf and Klaus M\o lmer}
\affiliation{
Lundbeck Foundation Theoretical Center for Quantum System Research, Department of
Physics and Astronomy, University of Aarhus, DK-8000 Aarhus C, Denmark.}

\date{\today}

\begin{abstract}
A protocol is provided to reconstruct the Wigner function for the motional state of a trapped ion via fluorescence detection
on another ion in the same trap.  This ``sympathetic tomography" of a dark ion without optical transitions suitable for state measurements is based on the mapping of its motional state onto one of the collective modes of the ion pair. The quantum state of this vibrational eigenmode is subsequently measured through sideband excitation of the bright ion. Physical processes to implement the desired state transfer and read-out are derived, and the accomplishment of the scheme for different mass ratios is evaluated.


\end{abstract}

\maketitle

\section{Introduction}

In quantum physics, states are represented by a wave function or a density matrix, which may in turn be represented as a phase space quasi-probability distribution. The Wigner function provides such a representation of the motional state of a particle, and it is convenient both because it provides a useful visualization of the position and momentum contents of the state \cite{wigner} and because it reveals non-classical properties of the system associated with quantum superposition states  \cite{nonclassicality}.

The ability to create and maintain superposition states is of relevance in quantum metrology and quantum information applications, and it is a key element in attempts to investigate the quantum-classical correspondence through experimental studies of larger and more complex physical systems \cite{penrose,arndt,cirac}. Experimental methods exist to reconstruct the motional state of a harmonically trapped ion by subjecting it to laser fields and coupling its internal and motional degrees of freedom \cite{wineland-sq,wineland}. For particles with no known or useful optical transitions, however, reconstruction of the motional state would naturally proceed via  monitoring of its spatial distribution at different times after its release from the trap - a formidable task as it requires precise elimination of any fields that would perturb the ion motion and distort the imaging.

In this paper we suggest to determine the motional quantum state of a trapped charged particle by use of an atomic ion in the same trap on which it is possible to perform laser excitation and fluorescence detection. We will refer to these particles as particle 1 and 2 and as the dark and bright ion in the following. The motion of the two ions is coupled via their Coulomb interaction, and our proposal is related to the idea of sympathetic cooling, where laser cooled ions exchange energy with other ions by their Coulomb interaction and thus bring them into thermal equilibrium at low temperature. 

The outline of the paper is as follows. In Sec. II, we determine the eigenmodes of the trapped ions and how they are related to the motion of the individual particles. In Sec. III, we discuss how the Wigner function for the dark particle motion is related to the complete Wigner function for the two particle problem, and we show how a physical operation on the system makes the dark ion motion available by measurements of only one of the collective modes. In Sec. IV we analyze the physical parameters for different mass rations between the bright and the dark ion. In Sec. V we present a conclusion and outlook.

\begin{figure}[t]
\centering {\includegraphics[width=2.5cm]{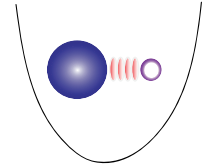}}
\caption{(Color online) Schematic setup for sympathetic tomography experiment. A dark and a bright charged particle are oscillating around their equilibrium positions in a harmonic one-dimensional ion trap.
}\label{fig:setup}
\end{figure}

\section{Trap eigenmodes}

In this section, we investigate the axial vibrational modes of a pair of coupled ions in a linear Paul trap. This problem has been studied before, but unlike the conventional diagonalization of the potential in the Lagrangian formulation, \cite{james,morigi}, we shall base our analysis on a diagonalization of the Hamiltonian into a pair of uncoupled oscillators. The eigenfrequencies and the  precise form of the modes are equivalent in the Lagrangian and Hamiltonian approaches, but the concrete steps in our diagonalization of the Hamiltonian will become useful in the following.

We assume that the ions are confined to a regime so that the trapping potential is well described by a quadratic expression in their excursions around their equilibrium position, and that they are transversally well confined, so that we can consider only the motion along the $x$-axis of a harmonic trap.

The total Hamiltonian is then given by,
\begin{equation}\label{v}
{H}=\frac{{p}_1^2}{2m_1}+\frac{{p}_2^2}{2m_2}+
\frac{1}{2}k{q}_1^2+\frac{1}{2}k{q}_2^2+\frac{e^2/(4\pi\epsilon_0)}{d_{0}-q_{1}+{q}_2},
\end{equation}
where $m_i$ is the mass, and ${q}_i(p_i)$ is the position (momentum) of the $i^{th}$ ion relative to its classical equilibrium position $q_i=x_i-x_{i,eq}$, ($i=1,2$). In our work we consider the situation where the dark ion is heavier than the bright ion, $m_1\geq m_2$. The internal state structure of ion 1 is not specified, but we assume that ion 2 has a two-level internal structure that can be excited by resonant laser fields.

In (\ref{v}), $k$ denotes the spring constant of harmonic motion along the $x$-axis in the trap, related to the oscillation frequencies of a single ion of any of the two species by $k=m_{i}\omega_{i}^2$.

The equilibrium distance between two ions of charge $e$ is $d_0=(\frac{2e^2}{4\pi\epsilon_{0}k})^{1/3}$ \cite{equ}, and
expanding the Coulomb interaction potential to second order the
Hamiltonian of the system is given by,
\begin{equation}\label{QH}
\hat{H}=\frac{\hat{p}_1^2}{2m_1}+\frac{\hat{p}_2^2}{2m_2}+
\frac{1}{2}m_1\omega_{1}^2\hat{q}_1^2+\frac{1}{2}m_{2}\omega_{2}^2\hat{q}_2^2+
\frac{1}{2}k(\hat{q}_2-\hat{q}_1)^2,
\end{equation}

where we use symbols $\hat{q}_i$ and $\hat{p}_i$ to recall that the positions and momenta are operators, which fulfil the canonical communication relations, i.e. $[\hat{q}_i,\hat{p}_j]=i\hbar\delta_{ij}$.
%In addition, $\omega_{i}$ denotes the frequency of the vibrations of the ion of mass $i$ in the trap.
The Hamiltonian (\ref{QH}) has the form of two coupled harmonic oscillators, and it can conveniently be put in matrix form,
\begin{eqnarray}
\hat{H}=X^{T}\textbf{H}X,
\end{eqnarray}

where,
\begin{eqnarray}\label{CH}
%\textbf{H}&=&
\textbf{H}&=&
\left( {\begin{array}{cccc}
 m_1\omega_{1}^2 & -k/2 & 0 & 0 \\
 -k/2 & m_2\omega_2 & 0 & 0  \\
 0 & 0 & 1/(2m_1) & 0 \\
 0 & 0 & 0 & 1/(2m_2)\\
 \end{array} } \right),\\
   X^{T}&=&
 \left( {\begin{array}{cc}
 \hat{q}_1 , \hat{q}_2 , \hat{p}_1 , \hat{p}_2
\end{array} } \right).
\end{eqnarray}

We aim to find the transformation of the variables,
\begin{eqnarray}\label{m}
\tilde{X}=MX,
\end{eqnarray}
leading to a diagonal Hamiltonian coefficient matrix
\begin{eqnarray}
\tilde{\textbf{H}}=(M^{-1})^T\textbf{H}M^{-1}.
\end{eqnarray}

$M$ must be a  symplectic transform \cite{symp} in order to preserve the canonical commutator relations. While a linear mixing of the two oscillators may bring the potential terms represented in the upper left quadrant of $\tilde{\textbf{H}}$ in diagonal form, in order to avoid the introduction of momentum cross terms in the lower right quadrant, we must first bring the kinetic energy operator into a diagonal form.

\subsection{Squeezing transformation} 
 
For this purpose, we introduce the scale, or squeezing, transformation,
\begin{eqnarray}\label{squeezing}
\left( {\begin{array}{c}
 \hat{q}'_1 \\
 \hat{q}'_2 \\
 \hat{p}'_1 \\
 \hat{p}'_2 \\
 \end{array} } \right)
=
\left( {\begin{array}{cccc}
 s & 0 & 0 & 0 \\
 0 & s^{-1} & 0 & 0  \\
 0 & 0 & s^{-1} & 0 \\
 0 & 0 & 0 & s
 \end{array} } \right)
 \left( {\begin{array}{c}
 \hat{q}_1 \\
 \hat{q}_2 \\
 \hat{p}_1 \\
 \hat{p}_2 \\
 \end{array} } \right),
\end{eqnarray}
or equivalently,
\begin{eqnarray}\label{m1}
X'=M_1X,
\end{eqnarray}
where, $s=(\frac{m_1}{m_2})^{1/4}$ is the rescaling parameter. This idea is illustrated and explained geometrically in \cite{kimpaper}.

Under the  transformation Eq. (\ref{m1}), the Hamiltonian coefficient matrix takes the form,
\begin{eqnarray}
\textbf{H}'=
\left( {\begin{array}{cccc}
 \mu\omega_1^2 & -k/2 & 0 & 0 \\
 -k/2 & \mu\omega_2^2 & 0 & 0  \\
 0 & 0 & 1/(2\mu) & 0 \\
 0 & 0 & 0 & 1/(2\mu)
\end{array} } \right),
\end{eqnarray}

where $\mu=\sqrt{m_1m_2}$.

\subsection{Mode mixing transformation}

At this point, one may perform a linear mixing of the two oscillator modes and eliminate the coupling terms in the potential. Note that this mixing is equivalent to a beam splitter transformation acting on two quantized fields, and it is equivalent to a spatial rotation of a two dimensional oscillator, conveniently parametrized by a rotation angle $\varphi$,
\begin{eqnarray}\label{rotation}
\left( {\begin{array}{c}
 \hat{q}''_1 \\
 \hat{q}''_2 \\
 \hat{p}''_1 \\
 \hat{p}''_2 \\
 \end{array} } \right)
=
\left( {\begin{array}{cccc}
 \cos\varphi & \sin\varphi & 0 & 0 \\
 -\sin\varphi & \cos\varphi & 0 & 0  \\
 0 & 0 & \cos\varphi & \sin\varphi \\
 0 & 0 & -\sin\varphi & \cos\varphi
 \end{array} } \right)
 \left( {\begin{array}{c}
 \hat{q}'_1 \\
 \hat{q}'_2 \\
 \hat{p}'_1 \\
 \hat{p}'_2 \\
 \end{array} } \right),\nonumber \\
\end{eqnarray}
or in the compact form,
\begin{equation}\label{m2}
\tilde{X}=M_2X'.
\end{equation}

With the correct choice of rotation angle,
\begin{eqnarray}\label{tan}
\tan2\varphi=\frac{1}{\sqrt{\frac{m_1}{m_2}}-\sqrt{\frac{m_2}{m_1}}},
\end{eqnarray}

it is possible to put the Hamiltonian into the desired diagonal form,
\begin{eqnarray}\label{f}
\tilde{\textbf{H}}=
\left( {\begin{array}{cccc}
 \mu\omega_{-}^2/2 & 0 & 0 & 0 \\
 0 & \mu\omega_{+}^2/2 & 0 & 0  \\
 0 & 0 & 1/(2\mu) & 0 \\
 0 & 0 & 0 & 1/(2\mu)
 \end{array} } \right).
\end{eqnarray}

The eigenmode frequencies are given by,
\begin{eqnarray}\label{freq}
\omega_{\pm}^2=\omega_2^2\bigl(1+\frac{m_2}{m_1}\pm\sqrt{1+(\frac{m_2}{m_1})^2-\frac{m_2}{m_1}}\bigr),
\end{eqnarray}
in agreement with the analysis in \cite{morigi}, and they represent collective motion of both ions. For the well studied case of identical ions, $\omega_+/\omega_- = \sqrt{3}$, and the eigenmodes represent the relative motion and center of mass motion of the ions. For different masses, the modes given by our expressions are more complicated.  The prospect of storing quantum states in the motional degree of freedom of one or several ions has led to numerous recent experiments which demonstrate the motional coupling in systems of identical and different ions \cite{wineland-which,blatt}. The normal mode frequencies have thus been observed in experiments, both with trapped atomic ions such as $Mg^+$ and $Be^+$ \cite{fexp}, and as a very precise diagnostic tool, allowing the determination of the mass of a dark molecular ion via probing of the collective mode sideband excitation frequency of a bright atomic ion \cite{michael,schiller}.

The individual ion motion associated with the normal modes is identified by the final outcome of the squeezing
Eq. (\ref{m1}) and the rotation transformations Eq. (\ref{m2}), and is formally given by
\begin{equation}\label{m}
 M=M_2M_1.
\end{equation}

\section{Wigner function reconstruction}\label{III}

In the previous section we showed that the normal modes of two different ions in a linear Paul trap are obtained by squeezing and rotation transformations of the single particle position and momentum coordinates.

The complete Wigner function of both particles is a pseudo probability function $W(q_1,p_1,q_2,p_2)$ on the four dimensional phase space of the particles. Alternatively, the Wigner function can be represented as a function of complex arguments $W(\alpha_1,\alpha_2)$, where the real (imaginary) parts of $\alpha_i$ are proportional to $q_i$ ($p_i$). After convolution with a Gaussian, the Wigner function leads to the Husimi- or Q-function, which is the overlap of the state with the coherent (product) states $|\alpha_1,\alpha_2\rangle$.

In medical diagnostics, tomography uses a number of two-dimensional images recorded at different angles to reconstruct a three-dimensional image of a patient. Likewise, quantum state tomography reconstructs a two-dimensional phase space distribution by a mathematical transformation of sequentially measured one dimensional quadrature distributions. This method has been applied to traveling wave quantum fields \cite{traveling}, to the single mode field inside a cavity \cite{cavity}-\cite{zubairy}, and to neutral and charged particle motion \cite{skovsen}.

For trapped ions, the reconstruction of the motional state is most conveniently carried out by a different method \cite{wineland}, which transfers information about the motional quantum state onto the internal state of the ions, which is subsequently read out by fluorescence. If a trapped ion is excited with a laser on the lower sideband in the Lamb-Dicke regime, the Rabi frequency of excitation depends on the motional eigenstate via a coupling strength $\propto \sqrt{n}$, and therefore monitoring of the internal state probabilities as function of time permits reconstruction of the motional eigenstate populations $p_n$, \cite{wineland-sq}. The $|n=0\rangle$ vibrational ground state is equivalent to a zero amplitude coherent state, and its population $p_0$, thus represents the $Q$- function at the
 phase space origin. Displacement of the ion prior to detection permits detection of the $Q(\alpha)$ function at any desired location $\alpha$ in phase space, and, moreover, as shown in \cite{wineland}, determining the full set of populations $p_n$ for such displaced copies of a single quantum states leads, via a simple formula, directly to the Wigner function $W(\alpha)$. For a single trapped ion, one applies the excitations on the lower motional sideband, and with two ions, we can use the difference in  eigenfrequency to selectively address either of the eigenmodes and determine their separate Wigner functions $W(\alpha_+)$ and $W(\alpha_-)$. Note, however, that knowledge of $W(\alpha_+)$ and $W(\alpha_-)$ alone does not permit reconstruction of the full two-mode Wigner function, and also not of the Wigner function of the individual particles.

 \subsection{Mapping on eigenmodes}

Whether one measures quadrature distribution functions or one assesses the phase space distribution via displacements and sideband excitations, the Wigner function for two particles, two field modes or spatial motion in two dimensions, is a function of two complex or four real variables. The sampling of such a function is thus a very elaborate task. Fortunately, it is not needed to acquire information of the full two-mode Wigner function and extract from this quantity the required single particle information. Instead, we propose to induce a physical process which maps the quantum state of particle 1 onto one of the eigenmodes, and subsequently perform single-mode tomography on that mode.

What we seek is a physical process which maps the particle 1 state onto the $(-)$ oscillator state. This is equivalent to requesting that the unitary time evolution operator $U(T)$ maps the particle 1 observables onto the similar observables for the $(-)$ mode. If the expansion of the Heisenberg picture operators $\hat{q}_-(T),\hat{p}_-(T)$ coincides with the expansion of $\hat{q}_{1}(0),\hat{p}_{1}(0)$ on the collective mode operators  $\hat{q}_\pm(0)$, $\hat{p}_\pm(0)$, measurements on the collective $(-)$ mode after the unitary mapping yields precisely the same outcome as one would obtain for the quadrature measurements on particle 1 in the initial state. The previous section showed explicitly that the eigenmode linear combinations, indeed, are derived from the action of two physical operations: squeezing and  mixing/rotation on the system, and we will now detail, which physical operations, applied to the eigenmodes, will yield the desired mode mapping.

The eigenmode operators and the individual particle operators are related by the specific linear combinations expressed by the matrix $M$ in Eq.(\ref{m}), with $M_1$ and $M_2$ given in Eqs.(8,11). Inverting, Eqs.(9,12), we can therefore write the vector of individual particle quadrature operators in terms of the eigenmode operators
\begin{equation}
\label{inversemat}
X=(M_2M_1)^{-1}X'' = M_1^{-1}M_2^{-1} X''.
\end{equation}
The derivation of the eigenmodes in Sec. II was not assuming any actual implementation of the squeezing and mixing/rotation operations, but we  note that by actually performing such operations, the  eigenmode operators evolve into, e.g.,
\begin{equation}
\hat{q}_-(T) = U_1^\dagger U_2^\dagger \hat{q}_-(0) U_2 U_1,
\end{equation}
which is, indeed, given by the expression (\ref{inversemat}), if $U_1$ squeezes $\hat{q}''_1=\hat{q}_{-}$ and $\hat{p}''_2=\hat{p}_{+}$ by $1/s$ and their conjugate variables by $s$ (inverse of Eq.(8)), and $U_2$ performs a rotation among the eigenmodes by the angle $-\varphi$ (inverse of Eq.(11)).

In the Schr\"odinger picture, we wish to measure the time independent $\hat{q}_-,\hat{p}_-$ quadrature operators in the time evolved state of the system $|\Psi(T)\rangle = U_2 U_1|\Psi(0)\rangle$, or $\rho(T)=U_2 U_1 \rho(0) U_1^\dagger U_2^\dagger$, evolved first by the squeezing operation $U_1$ on the eigenmodes and then by the rotation operation $U_2$ among the eigenmodes of the system.

\subsection{Squeezing as a physical transformation}

\begin{figure}[t]%\label{ldplot}
\centering {\includegraphics[width=6.7 cm]{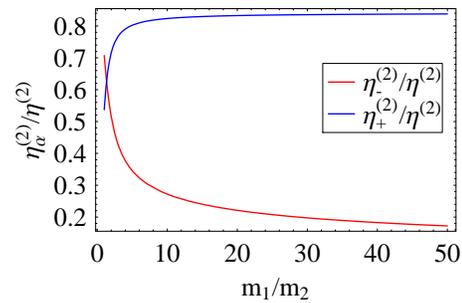}}
\caption{(Color online) Lamb-Dicke parameters $\eta_{\alpha}^{(2)}$ for the coupling of the bright ion 2 to the eigenmodes. The Lamb-Dicke parameters are shown in units of the Lamb-Dicke parameter $\eta^{(2)}$ for a single ion with mass $m_2$ as function of the  mass ratio $m_1/m_2$.}
\label{fig:ldplot}
\end{figure}

In order to squeeze the collective modes of motion, we propose to follow the scheme in Ref. \cite{wineland-sq} for motional squeezing of a single ion, by laser excitation detuned by twice the trap frequency. We address both collective modes via the bright ion 2, and we use two classical travelling waves with positive frequency parts with the same amplitudes and wave vectors $\kappa$ (or wave vector difference in case of Raman processes), but with different frequencies and phases, $E_{I}^{(+)}=E_{0} e^{i(\kappa\hat{q}_2-\omega_{I}t-\phi_{I})}$ and $E_{II}^{(+)}=E_{0} e^{i(\kappa\hat{q}_2-\omega_{II}t-\phi_{II})}$. The plane wave fields refer directly to the position operator $\hat{q}_2$ of the illuminated ion. In order to squeeze the $\alpha=\pm$ modes of collective motion, the frequencies  $\omega_{I}=\omega_{eg}+ 2\omega_{\alpha}$ and $\omega_{II}=\omega_{eg}- 2\omega_{\alpha}$ are chosen, where $\omega_{eg}$ is the frequency difference between the internal energy levels of the bright ion. Also, we put $\phi_{I}=-\phi_{II}=-\pi/2$. In the Lamb-Dicke regime, this effectively implements the Hamiltonian
\begin{eqnarray}\label{h_sq}
\hat{H}_{\alpha}=i\hbar\Omega_s[\eta_{\alpha}^{(2)}]^{2}(\hat{a}^{\dagger 2}_{\alpha}-\hat{a}^2_{\alpha})\hat{\sigma}_{x2},
\end{eqnarray}
where the Lamb-Dicke parameters $\eta_\alpha^{(2)}$ for the coupling of ion 2 to the $(-)$ and $(+)$ modes depend on the normal mode frequencies and on the masses of the two ions. Their values are provided in the Appendix, and they are shown in units of the Lamb-Dicke parameter $\eta^{(2)}$, for a single trapped ion with mass $m_2$ in
Fig. (\ref{fig:ldplot}) as functions of the mass ratio $m_1/m_2$. As the mass ratio of the particles increases, the Lamb-Dicke parameter for the coupling of the bright ion to the high frequency mode $\eta_+^{(2)}$ increases towards $0.84\eta^{(2)}$, while its coupling to the lower frequency mode decreases towards zero.

If the ion is prepared in the internal eigenstate $(|g\rangle +|e\rangle)/\sqrt{2}$ of the $\hat{\sigma}_{x2}$ Pauli operator, Eq.(19) is a squeezing Hamiltonian for mode $\alpha$, and we obtain the squeezing unitary operator of the mode $\alpha$,
\begin{eqnarray}
\hat{U}_1=e^{\xi\hat{a}^{2\dagger}_{\alpha}-\xi^*\hat{a}_{\alpha}^2}.
\end{eqnarray}

$\hat{U}_1$ provides the desired squeezing factor $s$ when $\xi\equiv\Omega_s[\eta_{\alpha}^{(2)}]^2t = \frac{1}{2}\ln{s}$. Therefore, to squeeze the collective modes by the desired amounts, one has to adjust the interaction times such that, $t_{\alpha}={\ln s}/{(2\Omega_{\alpha}[\eta_{\alpha}^{(2)}]^2)}$.
The two upper (red and blue) curves in Fig. (\ref{fig:plot2}) show the dependence of the squeezing duration $t_{\alpha}$ on the mass ratio in units of the single ion 2 squeezing time scale $([\eta^{(2)}]^2\Omega)^{-1}$.
Note that for equal masses, no squeezing is required.

\subsection{Mixing/rotation as a physical operation}

After the squeezing has been applied to both motional modes, we need to perform a rotation among the modes. This rotation is equivalent to a beam-splitter or mode-mixing transformation in optics \cite{su2}, but the modes have different frequencies, so a simple constant coupling will not suffice. There exist, however, proposals in cavity QED \cite{gao} to implement a mode mixing operation through the interaction of both modes with a single driven two-level system, and we shall adapt that scheme and illuminate the bright ion 2 with two classical standing wave lasers with positive frequency parts, $E_{I}^{(+)}=E_{0}\cos{(\kappa{q}_{2}+\theta_{I})}e^{-i(\omega_{I}t+\phi_{I})}$ and
$E_{II}^{(+)}=E_{0}\cos{(\kappa{q}_{2}+\theta_{II})}e^{-i(\omega_{II}t+\phi_{II})}$. Here, both lasers have the same amplitudes $E_{0}$ and wave numbers $\kappa$, but different frequencies $\omega_{I}$ and $\omega_{II}$, which will enable the resonant transfer of excitation among the vibrational modes via the ion 2 internal state.
We assume that the equilibrium position of the bright ion 2 is located at the anti-node of both standing waves, i.e. $\theta_{I}=\theta_{II}=0$. Under the above condition, the
interaction Hamiltonian between the bright ion and standing wave lasers is given by,
\begin{eqnarray}
\hat{H}&=&\frac{\hbar\Omega}{2}\hat{\sigma}_{+2}[(e^{-i\omega_{I}t-i\phi_{I}}+e^{-i\omega_{II}t-i\phi_{II}})\cos{(\kappa\hat{q}_{2})}\nonumber \\
&+&h.c.],\label{hint}
\end{eqnarray}
where, $\Omega$ is the coupling Rabi frequency and $\hat{\sigma}_{+2}=\left|e\right\rangle\left\langle g\right|$ is the raising operator of the internal ion state. The two laser frequencies are now chosen such that they bridge the frequency difference between the eigenmodes, $\omega_{I(II)}=\omega_{eg}+(-)(\omega_{+}-\omega_{-})$.

In the Lamb-Dicke regime, one may expand the interaction Hamiltonian of Eq. (\ref{hint}) up to second order in $\hat{q}_2$, and pass to the rotating frame given by the laser frequencies. Choosing $\phi_{I}=-\phi_{II}=\pi/2$, we obtain \cite{gao},
\begin{eqnarray}\label{hintint}
\hat{H}={i}\hbar\Omega\eta_{-}^{(2)}\eta_{+}^{(2)}(\hat{a}_{-}\hat{a}_{+}^{\dagger}-\hat{a}_{-}^{\dagger}\hat{a}_{+})\hat{\sigma}_{x2},
\end{eqnarray}

\begin{figure}[t]%\label{plot2}
\centering {\includegraphics[width=6.7cm]{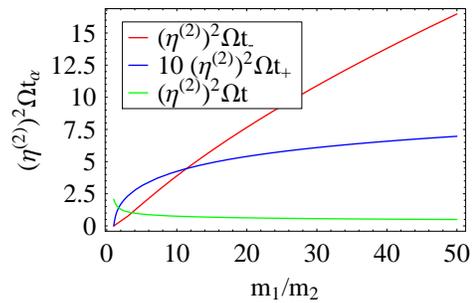}}
\caption{(Color online) Duration of the mixing (green) and squeezing operations (red and blue) as functions of $m_1/m_2$. The durations are given in units of the single ion time scale $([\eta^{(2)}]^2\Omega)^{-1}$. Note that the duration of the squeezing of the high-frequency mode is multiplied by 10 for improved visibility.}
\label{fig:plot2}
\end{figure}

Note that similar to our construction of the squeezing operators, we must prepare the ion in the internal superposition state $|g\rangle + |e\rangle)/\sqrt{2}$ to obtain the desired Hamiltonian action on the oscillator modes, which in this case causes a perfect unitary mixing or rotation among the modes,
\begin{eqnarray}
\hat{U}(\varphi)=e^{\varphi(\hat{a}_{-}\hat{a}_{+}^{\dagger}-\hat{a}_{-}^{\dagger}\hat{a}_{+})},
\end{eqnarray}
where, $\varphi=\Omega\eta_{-}^{(2)}\eta_{+}^{(2)}\hat{\sigma}_{x2}t$ defines the rotation angle.
To obtain the desired rotation angle $\varphi$, given by Eq.(13), one has to adjust the interaction time such that $\eta_{+}^{(2)}\eta_{-}^{(2)}\Omega t=\varphi$. The lower curve (green) in Fig. (\ref{fig:plot2}) shows this time as a function of the mass ratio $m_1/m_2$ in units of the single ion time scale $([\eta^{(2)}]^2 \Omega)^{-1}$, indicating that the duration of the rotation operation decreases with increasing mass ratio.

\subsection{Time scales}

\begin{figure}[t]%\label{plot2}
\centering {\includegraphics[width=6.7cm]{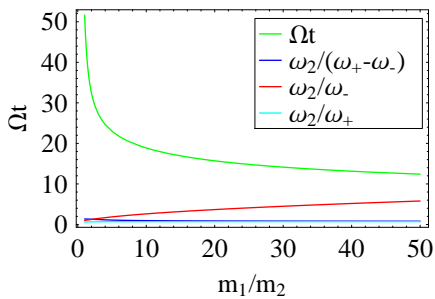}}
\caption{(Color online) Duration of mapping operations and fulfilment of sideband resolution. The upper (green) curve shows the duration of the mapping process in units of the single ion time unit $1/\Omega$ for $\eta_2^{(2)}=0.2$ vs. the mass ratio $m_1/m_2$. The lower curves show  $\omega_2/(\omega_+-\omega_-)$, $\omega_2/\omega_-$ and $\omega_2/\omega_+$.
As long as $\Omega < \omega_2$ the mapping is slow enough that the sidebands are well resolved.}
\label{fig:time}
\end{figure}

In summary, we have presented physical operations that map the motional state of ion 1 onto the eigenmode $(-)$, and we have shown that these operations can be carried out on time scales which are similar to the ones already demonstrated for the manipulation of single trapped ions. Since we selectively address the eigenmode sidebands, they must be resolved from the carrier and from each other, giving rise to the further requirement that the operation time must be sufficiently larger than $(\omega_+-\omega_-)^{-1}$, $\omega_+^{-1}$, and $\omega_-^{-1}$. In Fig. (\ref{fig:time}) we show the interaction time for the mode mixing process in units of the inverse Rabi frequency $1/\Omega$ (upper green curve) and compare it with $(\omega_+-\omega_-)^{-1}$, $\omega_+^{-1}$, and $\omega_-^{-1}$ (lower curves). Note that the latter are shown in units of $1/\omega_2$, the inverse trapping frequency, and that long interaction times and resolved sidebands are guaranteed for both the squeezing and mixing processes if $\Omega < \omega_2$, which will quite generally be the case.

In the beginning of this section we described how the Wigner function of the motional eigenmode can be determined by fluorescence measurements on the internal state of the  bright ion, after displacement operations applied to the eigenmode and excitation on the lower sideband transition. The diplacements may be done using the excitation mechanism provided in Ref. \cite{wineland}, and the fluorescence detection is efficiently carried out by means of the shelving method used also in \cite{wineland}.

We note that the eigenmodes are not mixed by the Coulomb and trapping potential, and therefore the tomographic measurement on the $(-)$ mode can be carried out in the interaction picture moving frame of that oscillator with adequate time to perform measurements on the internal state of the bright ion. In practical applications, however, the measurement time must of course be  short compared to heating and decoherence time scales in the system. In the limit of heavy dark ions, the low frequency of the $(-)$ mode might make it more susceptible to such decoherence, and one might instead transfer the state of particle 1 to the $(+)$ mode with the higher frequency. This requires a larger rotation angle $\varphi$, and hence the mapping process takes longer, suggesting that a criterion can be established for the optimum collective destination mode for the dark particle quantum state.

\section{Analysis}

In this section we present an overview of the quantitative estimates for our protocol in three different cases with $Be^+$ as the bright ion.

\subsection{Two ions with equal mass}
Consider the situation in which two \textit{Be$^+$} ions are located in the linear Paul trap. Making use of Eqs. (\ref{pq}), the normal motional modes of the system are given by,
\begin{eqnarray}\label{qemass}
\hat{q}_{-}&=&0.70\hat{q}_1+0.70\hat{q}_2,\nonumber \\
\hat{q}_{+}&=&-0.70\hat{q}_1+0.70\hat{q}_2,
\end{eqnarray}
with corresponding normal frequencies, $\omega_{-}=\omega_2$ and $\omega_{+}=\sqrt{3}\omega_2$ where $\omega_2$ refers to the oscillation frequency of only ion 2 in the trap. $\omega_{-}$ and $\omega_{+}$ correspond to the center-of-mass and stretch mode frequencies, and the normal modes are obtained by a mixing/rotation transformation with $\varphi=\pi/4$, c.f.,  Eq. (\ref{tan}). In this case, the Lamb-Dicke parameters are evaluated as $\eta_{-}^{(2)}=0.70\eta^{(2)}$ and $\eta_{+}^{(2)}=0.54\eta^{(2)}$ where $\eta^{(2)}$ is the Lamb-Dicke parameter of a single $Be^+$ ion located in the trap.

\subsection{A $Be^+$ and an $Al^+$ ion}

When a \textit{Be}$^+$ ion is coupled to an $\textit{Al}^+$ ion, ($m_1/m_2=3$) and the normalized eigenmodes of motion are given by,
\begin{eqnarray}
\hat{q}_{-}&=&0.98\hat{q}_1+0.21\hat{q}_2,\nonumber \\
\hat{q}_{+}&=&-0.54\hat{q}_1+0.84\hat{q}_2,
\end{eqnarray}
with corresponding eigenfrequencies, $\omega_{-}=0.67\omega_2$ and $\omega_{+}=1.49\omega_2$. The low frequency in-phase mode of motion has a larger share of the heavier particle's displacement, while particle $2$ contributes a larger amount of the high frequency mode as witnessed by the Lamb-Dicke parameters, $\eta_{-}^{(2)}=0.43\eta^{(2)}$ and $\eta_{+}^{(2)}=0.77\eta^{(2)}$.

In comparison to the case of equal masses, the mixing angle $\varphi=0.36$ is decreased and as a consequence the duration of the mixing process gets shorter.

\subsection{Larger mass differences}

 If the $Be^+$ ion, is coupled to a $20$ times heavier particle in the trap, the collective normal modes correspond to,
\begin{eqnarray}
\hat{q}_{-}&=&0.99\hat{q}_1+0.02\hat{q}_2,\nonumber \\
\hat{q}_{+}&=&-0.46\hat{q}_1+0.89\hat{q}_2,
\end{eqnarray}
with, $\omega_{-}=0.27\omega_{2}$ and $\omega_+=1.42\omega_{2}$, and $\eta_{-}^{(2)}=0.22\eta^{(2)}$ and $\eta_{+}^{(2)}=0.83\eta^{(2)}$. The mixing angle $\varphi=0.12$, and by increasing the ratio $m_1/m_2$ even further, the lower frequency converges to zero while the higher one tends to $\sqrt2\omega_2$. 

\begin{figure}[t]%\label{plot2}
\centering {\includegraphics[width=5.7cm]{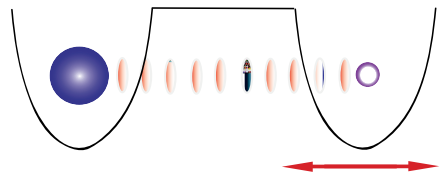}}
\caption{(Color online) Schematic setup for experiments with two ions trapped in separate local traps with harmonic oscillator shape near the trapping minima. As the trap holding the bright ion is brought closer to the dark particle, the individual particle modes mix, and eigenmodes are formed. Using controlled displacement of the bright ion trap, squeezing of the eigenmodes, and mixing of the modes with bichromatic fields, the motional state of an initially isolated dark particle may be transferred to a collective mode and read out by sideband excitation.}
\label{fig:set-up 2}
\end{figure}

\section{Conclusion}

We have in this article proposed a method to reconstruct the Wigner function for the quantum state of motion of a charged particle located in a linear ion trap, by transfer of the state to one of the eigenmodes shared with another ion in the trap on which laser excitation and fluoresecence detection is possible. The mapping of the quantum state is carried out by squeezing and mode mixing transformations which can all be implemented via illumination of the bright ion, and we showed that the operations are not more complex or time consuming than similar operations routinely applied to single trapped ions.

The coupled motion of trapped ions has been studied for a long time as a key element in applications for entanglement generation and quantum computing \cite{cz-gate,ms-gate,duanmonroe-gate}, where the motional state serves as a communication channel between the ion internal states. Recent investigations on the manipulation of the quantum states of motion of several particles \cite{wineland-which,blatt,fexp} show great potential for more delicate use of the motional degree of freedom, and we believe that the present work contributes to this new interesting research field. In particular, the demonstration, following \cite{gao}, that non-degenerate collective modes can be effectively coupled by absorption and stimulated emission of fields with the appropriately chosen frequencies, may have much wider applications than indicated here.

Our example dealt for simplicity with two ions located in the same trap, but with modern segmented surface electrode traps \cite{surfacetrap}, it is possible to prepare situations like the one depicted in Fig. 5, where two ions are located in separate harmonic potential minima. If the traps are far apart, the individual particle modes are not coupled, while for shorter distance, one can identify the eigenmodes and the associated squeezing and mixing transformations in the same manner as above. With time dependent potentials, it is also possible to move the traps, and bring the bright probe ion closer to the initially isolated dark particle. The oscillators with time dependent coupling are still governed by linear (symplectic) transformations of the quadrature observables and we can thus identify and carry out the physical transformations to map the quantum states for sympathetic tomography as we did in the single trap.
The Bloch-Messiah reduction \cite{braunstein} ensures that the desired transformation of the observables can always be obtained by a combination of squeezing and mode mixing operations.

Finally, we note the increasing interest in hybrid systems, where, e.g., trapped ions may interact with neutral atoms \cite{atom} or molecules \cite{polar}, cold quantum gases \cite{zipkes}, and optical cavity fields \cite{dantan}. The interactions are different in these cases, but similar methods may exist for sympathetic cooling and sympathetic tomography between the quantum states of the ion and of the partner system.

Discussions with Michael Drewsen and Uffe V. Poulsen and financial support from the European Union Integrated Project
AQUTE are gratefully acknowledged.

\appendix
\section{}

\subsubsection{Lamb-Dicke parameters}
In order to identify the Lamb-Dicke parameters, one has to find the explicit form of the normal modes. After applying rotation Eq. (\ref{squeezing}) and squeezing (\ref{rotation}), we have,
\begin{eqnarray}\label{pq}
\hat{q}_{-}&=&s \cos{\varphi} \hat{q}_{1}+s^{-1}\sin{\varphi}\hat{q}_2,\nonumber\\
\hat{q}_{+}&=&-s \sin{\varphi} \hat{q}_{1}+s^{-1}\cos{\varphi}\hat{q}_2,
%\end{eqnarray}
%\hat{p}_{-}&=&s^{-1}\cos{\varphi}\hat{p}_{1}+s\sin{\varphi}\hat{p}_2,\nonumber\\
%\hat{p}_{+}&=&-s^{-1}\sin{\varphi}\hat{p}_{1}+s\cos{\varphi}\hat{p}_2,
\end{eqnarray}
where, $s=\sqrt[4]{m_1/m_2}$ and $\varphi$ is defined by Eq. (\ref{tan}). Inverting the above equations, we obtain
\begin{eqnarray}\label{q2}
\hat{q}_2=s \sin{\varphi} \hat{q}_-+s \cos{\varphi} \hat{q}_{+}.
\end{eqnarray}
Laser light couples to the internal
degrees of freedom of ion 2 and thereby to the external degrees of
freedom of the collective motion. In the dipole approximation, the coupling involves the plane wave factor $e^{i\kappa \hat{q}_2}$, where $\kappa$ is the wave number of the light field (or the wave number difference in case of a two-photon Raman process) and $\hat{q}_2$ is the position operator of the resonantly excited particle. Substitution of $\hat{q}_2$ from Eq. (\ref{q2}) into the plane wave factor, and expansion of the collective mode position operators on rasing and lowering operators,
$\hat{q}_\alpha = \sqrt{\frac{\hbar}{2\mu\omega_{\alpha}}}(\hat{a}_{\alpha}+\hat{a}_{\alpha}^{\dagger})$ leads to the indentification of the collective mode Lamb-Dicke factors,
\begin{eqnarray}
\eta_{-}^{(2)}&=&\kappa\sin{\varphi}\sqrt{\frac{\hbar}{2m_2\omega_-}},\nonumber\\
\eta_{+}^{(2)}&=&\kappa\cos{\varphi\sqrt{\frac{\hbar}{2m_2\omega_+}}},
\end{eqnarray}
where the rotation angle is given by Eq.(13), and the mode eigenfrequencies are given by Eq.(15) in the text.

\end{document}